\documentclass[fleqn, usenatbib]{mnras}

\usepackage[T1]{fontenc}

\DeclareRobustCommand{\VAN}[3]{#2}
\let\VANthebibliography\thebibliography
\def\thebibliography{\DeclareRobustCommand{\VAN}[3]{##3}\VANthebibliography}

\usepackage{booktabs}
\usepackage{siunitx}
\usepackage{graphicx}
\usepackage{amsmath}
\usepackage{amssymb}
\usepackage{subcaption}
\usepackage{hyperref}
\usepackage{lipsum}
\usepackage{anyfontsize}

\usepackage{pgfplotstable}
\pgfplotsset{compat=1.18}
\pgfplotstableset{
precision=4,
}

\newcommand{\msol}{$\mathrm{M_{\sun}}$}

\newcommand{\cm}{$\mathrm{cm}$}
\newcommand{\barye}{$\mathrm{barye}$}
\newcommand{\K}{$\mathrm{K}$}
\newcommand{\s}{$\mathrm{s}$}
\newcommand{\erg}{$\mathrm{erg}$}
\newcommand{\gcmcubed}{$\mathrm{g}\,\mathrm{cm}^{-3}$}
\newcommand{\kmpersec}{$\mathrm{km}\,\mathrm{s}^{-1}$}

\title[.Ia Supernova Detonations]{Evidence of .Ia Supernova Detonations in 3D Hydrodynamical Simulations of Double Degenerate Mergers} 

\author[U. P. Burmester et al.]{
Uri Pierre Burmester$^{1}$\thanks{E-mail: uri.burmester@anu.edu.au (UPB)},
Lilia Ferrario$^{1}$,
R\"udiger Pakmor$^{2}$, and
Ivo R.~Seitenzahl$^{3}$
\\
$^{1}$Mathematical Sciences Institute, Australian National University, Canberra ACT 0200, AU\\
$^{2}$Max Planck Institute for Astrophysics, Karl-Schwarzschild-Strasse 1, 85748 Garching, DE\\
$^{3}$Heidelberger Institut f\"ur Theoretische Studien (HITS), Schloss-Wolfsbrunnenweg 35, 69118 Heidelberg, DE\\
}

\date{Accepted XXX. Received YYY; in original form ZZZ}

\pubyear{2025}

\usepackage{newtxtext,newtxmath}

\begin{document}
\label{firstpage}
\pagerange{\pageref{firstpage}--\pageref{lastpage}}
\maketitle

\begin{abstract}
We report detailed 3D simulations of 1.1~\msol\ Oxygen-Neon (ONe) white dwarfs (WDs) merging with a 0.35~\msol\ helium WD, conducted with the moving-mesh hydrodynamic code \texttt{AREPO}. The simulations utilise self-consistent chemical profiles for the primary WD which were generated by a stellar evolution code incorporating the effects of semi-degenerate carbon burning. We find that a helium detonation is ignited at the base of the helium layer, starting a thermonuclear runaway which encircles the WD and ejects material as a sub-luminous supernovae. Our canonical simulation, (C-120), ejects 0.103~\msol\ of primarily $^{4}\mathrm{He}$, $^{28}\mathrm{Si}$, and $^{32}\mathrm{S}$, after which the primary begins accreting again from the surviving secondary. Our results depend qualitatively on the ``inspiral time'' simulation parameter, which describes the length of a period of accelerated angular momentum loss. For example, the binary does not survive when inspiral time is too long. We compare the results using our self-consistent chemical profiles to a constant-composition WD structure and find the same explosion pattern when the inspiral time is short. However, we are able to obtain a typical Type Ia supernova (SN\,Ia) which destroys the primary by using the constant-composition structure and long inspiral. The shock-convergence in this simulation follows the ``x-scissor mechanism'' described by Gronow et al. in 2020, and causes a secondary detonation due to higher temperatures and higher number density of $^{12}\mathrm{C}$ at the convergence site. These results highlight the potential for unrealistic outcomes when conducting simulations that incorporate unrealistically large enhancements in angular momentum losses and (or) non-realistic chemical structures for the primary WD.
\end{abstract}




\section{Introduction}

The interaction and merging of close binary systems containing two WDs are at the forefront of many discussions involving the astrophysics of transients. These include investigations into the origins and mechanisms behind SNe\,Ia \citep{ruiterTypeIaSupernova2020, liuTypeIaSupernova2023, ruiterTypeIaSupernova2025}, the processes leading to the generation of neutron stars through the accretion-induced collapse (AIC) or merger-induced collapse (MIC) of WDs \citep{ruiterFormationNeutronStars2019}, the formation of magnetars \citep{duncanFormationVeryStrongly1992}, the possible association with fast radio bursts \citep[FRBS,][]{lorimerBrightMillisecondRadio2007}, and the role of merging binary WDs as sources of low-frequency gravitational waves \citep{ruiterFormationNeutronStars2019}. The merging of two WDs have also been convincingly associated with the formation of ultra-massive, strongly magnetic WDs \citep{kawkaNonexplosiveStellarMerging2023}. Generating accurate simulations to investigate the outcomes of double WD mergers will thus give us crucial clues regarding some of the most intriguing transient events in the universe. 

Numerically, achieving a comprehensive and internally-consistent approach to modelling stellar mergers has been challenging due to physical variables that vary by dozens of orders of magnitude and must account for complex physics. This complexity arises from the involvement of numerous physical processes that extend across very large spatial and temporal scales. In spite of these problems, there has been gradual progress on the investigation of double degenerate mergers via exploration of the parameter space \citep[see e.g.][]{danStructureFateWhite2014}, studying results obtained through the use of different simulation codes \citep[e.g.][]{richardsonExploringViscosityEffects2024}, and comparing the outcomes of simulations with robust catalogues of observations \citep[e.g.][]{deZwickyTransientFacility2020}.

It is generally accepted that the thermonuclear explosions of carbon–oxygen (CO) WDs are responsible for SNe\,Ia events and the creation of heavy chemical elements \citep{hoyleNucleosynthesisSupernovae1960} -- crucial for enriching the interstellar medium and shaping the chemical evolution of galaxies \citep{eitnerObservationalConstraintsOrigin2020}. However, the precise nature of their progenitors, stellar masses, and ignition mechanisms are as yet largely unknown. Key questions include under which conditions a thermonuclear explosion occurs, whether this will transition into a detonation, and what effect our choice of numerical scheme has on the outcome \citep{seitenzahlSpontaneousInitiationDetonations2009, 
 seitenzahlInitiationDetonationGravitationally2009}. Various combinations of CO WD binaries are relatively well-explored in the literature \citep[see, among others,][]{burmesterAREPOWhiteDwarf2023, pakmorFateSecondaryWhite2022, gronowSNeIaDouble2020}, whereas much less attention has been paid to WDs with different chemical compositions. ONe\,WDs are interesting in part due to the much lower presence of carbon in their internal structures -- the higher Coulomb repulsion of $^{16}\mathrm{O}$ and $^{20}\mathrm{Ne}$ suggest a less explosive outcome.

In this study, we investigate the possibility of an ONe\,WD exploding and triggering a SN event, and, if an explosion is ignited, determine the type of SN event that might result. Some of the possible outcomes might include: (i) a Type Ia supernova (SN\,Ia), (ii) the formation of a more massive WD, (iii) the generation of a neutron star via the merger-induced collapse of the WD, or (iv) a supernova explosion of another type. In this context, the pioneering work of \citet{nomotoConditionsAccretioninducedCollapse1991} showed that the ultimate fate of an accreting ONe\,WD  depends on whether the release of nuclear energy is faster than electron capture behind the deflagration wave. The former is mostly determined by the speed of the wave, while the latter by the density. Thus, it is still far from clear whether such an event would lead to the formation of a neutron star or to a complete or partial thermonuclear explosion with the latter leaving behind a bound remnant such as a ONe\,WD with a sizeable fraction of Fe-group elements \citep[see the work of][]{jonesRemnantsEjectaThermonuclear2019}. The WD LP\,40-365 \citep{vennesUnusualWhiteDwarf2017} could be an example of such a remnant, although its radius seems to be too large for the models of ONeFe\,WDs derived by \citet{jonesRemnantsEjectaThermonuclear2019}. The hydrodynamic explosion simulations of ONe\,WDs for a range of masses below the Chandrasekhar's mass conducted by \citet{marquardtTypeIaSupernovae2015} showed that the synthetic ejecta structures and observables resembled those of their CO\,WD counterparts. Thus, they concluded that exploding ONe\,WDs, if ignited successfully, could potentially be progenitor systems of SNe\,Ia, albeit much rarer. 

Another potential explosion outcome are SNe of type ``.Ia''. \citet{shenThermonuclearIaSupernovae2010} provide a description of the outcome of the early evolution of an AM\,CVn system in which helium is accreted onto the WD primary under conditions favourable for unstable thermonuclear ignition. \citet{shenThermonuclearIaSupernovae2010} use hydrodynamic simulations to show that turbulent motions induced in the convective burning stage in the He envelope could produce a sub-luminous detonation event creating heavy $\alpha$-chain elements ($^{40}\mathrm{Ca}$ through $^{56}\mathrm{Ni}$) and unburnt helium. 

The present work is an extension of the 3D high-resolution hydrodynamical simulations of \citet{burmesterAREPOWhiteDwarf2023} (Paper\,I hereafter) generated with the moving-mesh code \texttt{AREPO}. The calculations presented in Paper\,I modelled the merging of a 1.1\,\msol\ CO\,WD with a 0.35\,\msol\ secondary He\,WD and resulted in a detonation at the base of the helium layer of the CO\,WD that then triggered an off-centre carbon detonation and the total disruption of the primary CO\,WD. The simulations showed that the WD companion survives the explosion with a leftover mass of about 0.22\msol\ and moving at a speed greater than 1\,700\,km\,s$^{-1}$, consistent with the observational findings of hypervelocity WDs \citep[e.g.,][]{shenThreeHypervelocityWhite2018, el-badryFastestStarsGalaxy2023}.

In this work, the CO\,WD is replaced with an ONe\,WD featuring a realistic chemical profile. The aim is to establish whether a detonation can still occur and whether exploding ONe\,WDs could serve as progenitor systems for SNe\,Ia. This experiment is primarily motivated by the fact that previous simulations of exploding ONe\,WDs were not triggered by interaction with a companion star but were artificially ignited.

The codes and the parameters used to set up the simulations are described in Section \ref{sec:methods}. We show the results of our calculations and the stability of the new structures in Section~\ref{sec:results}, the results are discussed in Section~\ref{sec:discussion} and a summary is provided in Section~\ref{sec:conclusions}.

\section{Methods} \label{sec:methods}

To run our merger simulations we have used the code \texttt{AREPO}, described in detail in \citet{springelPurSiMuove2010,pakmorImprovingConvergenceProperties2016,weinbergerAREPOPublicCode2020}. \texttt{AREPO} solves the ideal MHD equations using a second-order finite-volume method on a Voronoi mesh that moves with the flow and is generated at each timestep, thus ensuring a nearly Lagrangian behaviour. The flux across cell boundaries is calculated by employing a Harten-Lax-van Leer discontinuities (HLLD) approximate Riemann solver as described by \citet{pakmorMagnetohydrodynamicsUnstructuredMoving2011}. We note that while this is an MHD solver, the simulations presented in this work are purely hydrodynamic. Self-gravity is included with a one-sided octree solver. The gravitational force is softened with a softening length of $2.8$ times the radius of a cell to suppress non-genuine two-body interactions, but force the softening to be at least $10$\,km.

\begin{figure}
    \centering
    \includegraphics[width=\linewidth]{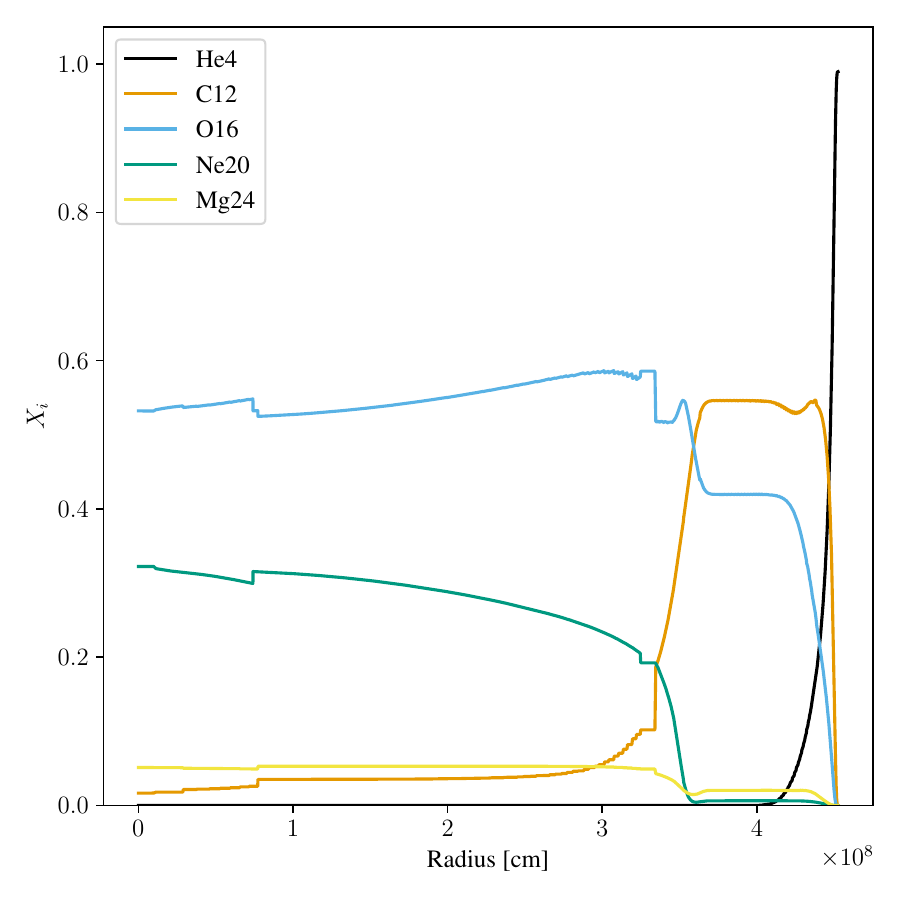}
    \caption{Radial chemical composition profile of the WD structures computed by C19. We show here the elements included in the initial 5-species network. Note that C19's original model included several other isotopes, such as Ne22 and Na23, which are not included in the initial network. Hence, the total is less than unity - the 3D equivalent to the 1D structure reapportioned the weighting to the five remaining species to normalise the composition.}
    \label{fig:maria_profile}
\end{figure}

Refinement and de-refinement are applied when a cell's mass differs from our target mass resolution ($10^{-6}$\,\msol), \texttt{AREPO} will attempt to split cells whose mass is over twice the target mass resolution and merge cells whose mass is half the target mass resolution. Further refinement is applied if the volume of a cell is ten times larger than its smallest immediate neighbour to prevent sharp gradients in cell volume. The maximum volume for cells is set to $10^{30}$\,cm$^3$ to stop derefinement of the background mesh. 

\begin{figure*}
    \centering
    \includegraphics[width=0.83\linewidth]{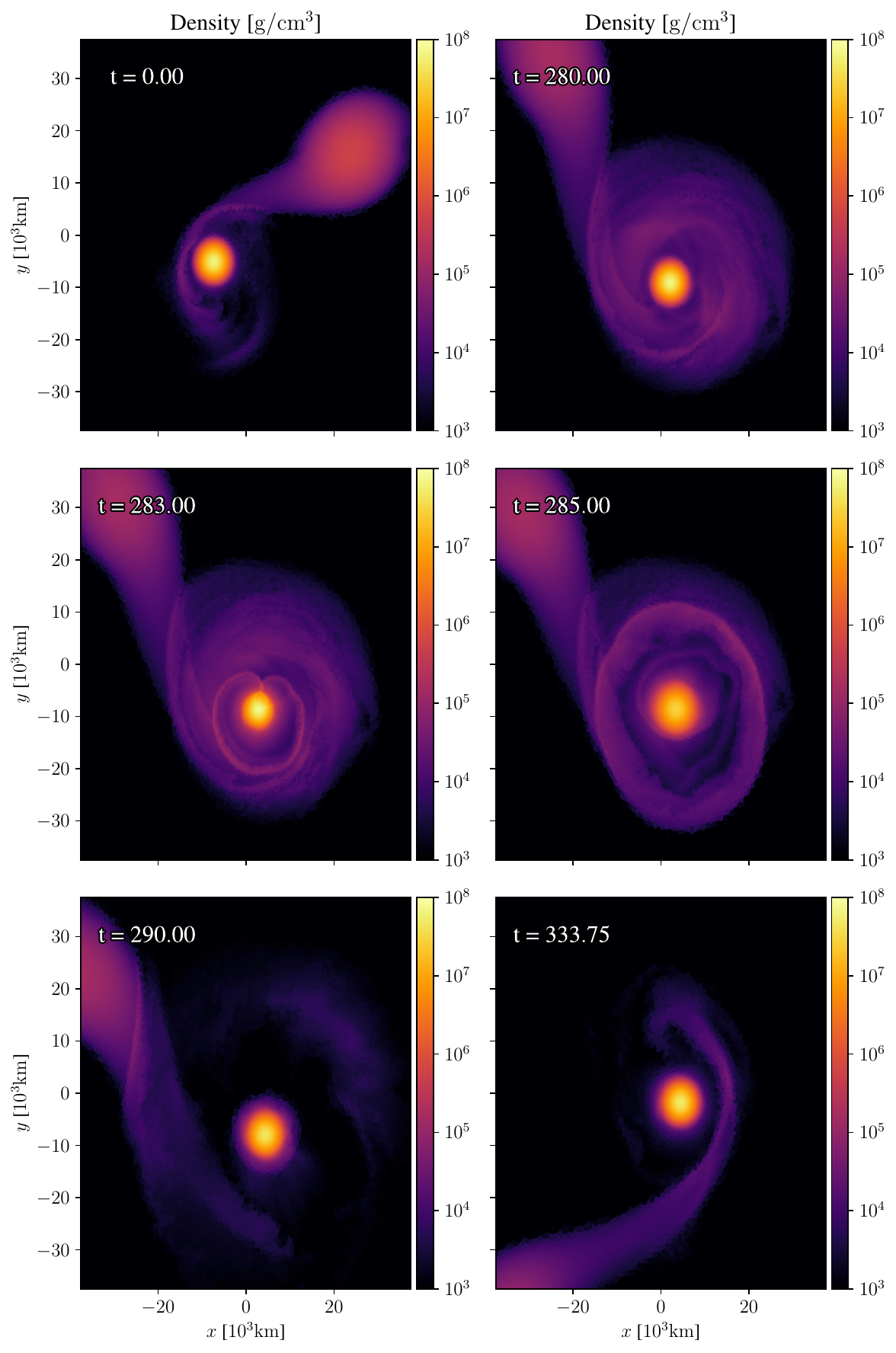}
    \caption{Evolution of density for the C-120 model. The time $t=0$ indicates the end of the inspiral period and the onset of significant mass transfer. A disturbance forms at the base of the helium layer which encircles the primary and ejects approximately $0.103$ solar mass of material from the binary. The location of the heavier elements in the last panel can be seen in Figure~\ref{fig:pcolor_late_time}.}
    \label{fig:evolution}
\end{figure*}

\begin{table}
\centering
\caption{The 55-species nuclear reaction network list which is coupled to \textsc{AREPO}, used in \citet{pakmorThermonuclearExplosionMassive2021}. Entries in bold face are those present in the 13-species network.}
\label{tab:55_13_network_comparison}
\resizebox{0.7\linewidth}{!}{
\begin{tabular}{|l|l|l|l|l|}
El.          & .             & .             & .             & .             \\ \hline
n            & O17           & Mg26          & \textbf{S32}  & Sc43          \\
p            & F18           & Al25          & S33           & \textbf{Ti44} \\
\textbf{He4} & Ne19          & Al26          & Cl33          & V47           \\
B11          & \textbf{Ne20} & Al27          & Cl34          & \textbf{Cr48} \\
\textbf{C12} & Ne21          & \textbf{Si28} & Cl35          & Mn51          \\
C13          & Ne22          & Si29          & \textbf{Ar36} & \textbf{Fe52} \\
N13          & Na22          & Si30          & Ar37          & Fe54          \\
N14          & Na23          & P29           & Ar38          & Co55          \\
N15          & Mg23          & P30           & Ar39          & \textbf{Ni56} \\
O15          & \textbf{Mg24} & P31           & K39           & Ni58          \\
\textbf{O16} & Mg25          & S31           & \textbf{Ca40} & Ni59         
\end{tabular}
}
\end{table}

To model the partially degenerate fluid we use the Helmholtz equation of state \cite[HES,][]{timmesAccuracyConsistencySpeed2000}. We also employ a nuclear reaction network which is fully coupled to \texttt{AREPO} \citep{pakmorThermonuclearExplosionMassive2021} and draws on reaction rates provided by Joint Institute for Nuclear Astrophysics \texttt{REACLIB} database\footnote{\url{http://groups.nscl.msu.edu/jina/reaclib/db/}}. We employ either the 13-species or the 55-species network (see Table~\ref{tab:55_13_network_comparison}) depending on the scenario. Typically, the 13-species network is used to simulate a long mass-transfer phase until a detonation or other interesting behaviour is observed. The simulation is then restarted using the 55-species network with the last 13-species snapshot before detonation onset used as input. The choice of network size is discussed further in Appendix~\ref{app:influence_network}.

\citet{burmesterAREPOWhiteDwarf2023} also compares the yields of the 55-species network with a 384-species network which is available in post-processing. The nuclear reaction network is operative for cells that have $T > 2 \times 10^7$\,\K. Shocked cells characterised by $\nabla \cdot \vec{v} < 0$ and $|\nabla P| r_{\rm cell} / P < 0.66 $, where $\vec{v}$ is the cell's velocity, $r_{\rm cell}$ is the cell's radius, and $P$ is the pressure have been excluded to assure numerical stability \citep[see][for further information]{seitenzahlSpontaneousInitiationDetonations2009,fryxellHydrodynamicsNuclearBurning1989}. Hydrogen is not present in the atmospheric model of the ONe\,WD primary because its total fractional mass is negligible and essentially impossible to resolve at our mass resolution, has negligible effects on the resulting chemical profile, and would make the simulations excessively time-consuming. 

Many simulations in the literature employ a constant composition, isothermal WD and a numerical integrator that uses equations of hydrostatic equilibrium to generate the pressure and interior mass distribution of the WD. Other variables, such as density, are then derived through a power-law relationship (i.e. a polytrope) or by using an equation of state (EOS). In this study, we use realistic chemical profiles for the WD primary and compare the results with those obtained using homogeneous composition WD models (see section \ref{sec:models}).

\subsection{Models} \label{sec:models}

\begin{figure*}
    \centering
    \includegraphics[width=\linewidth]{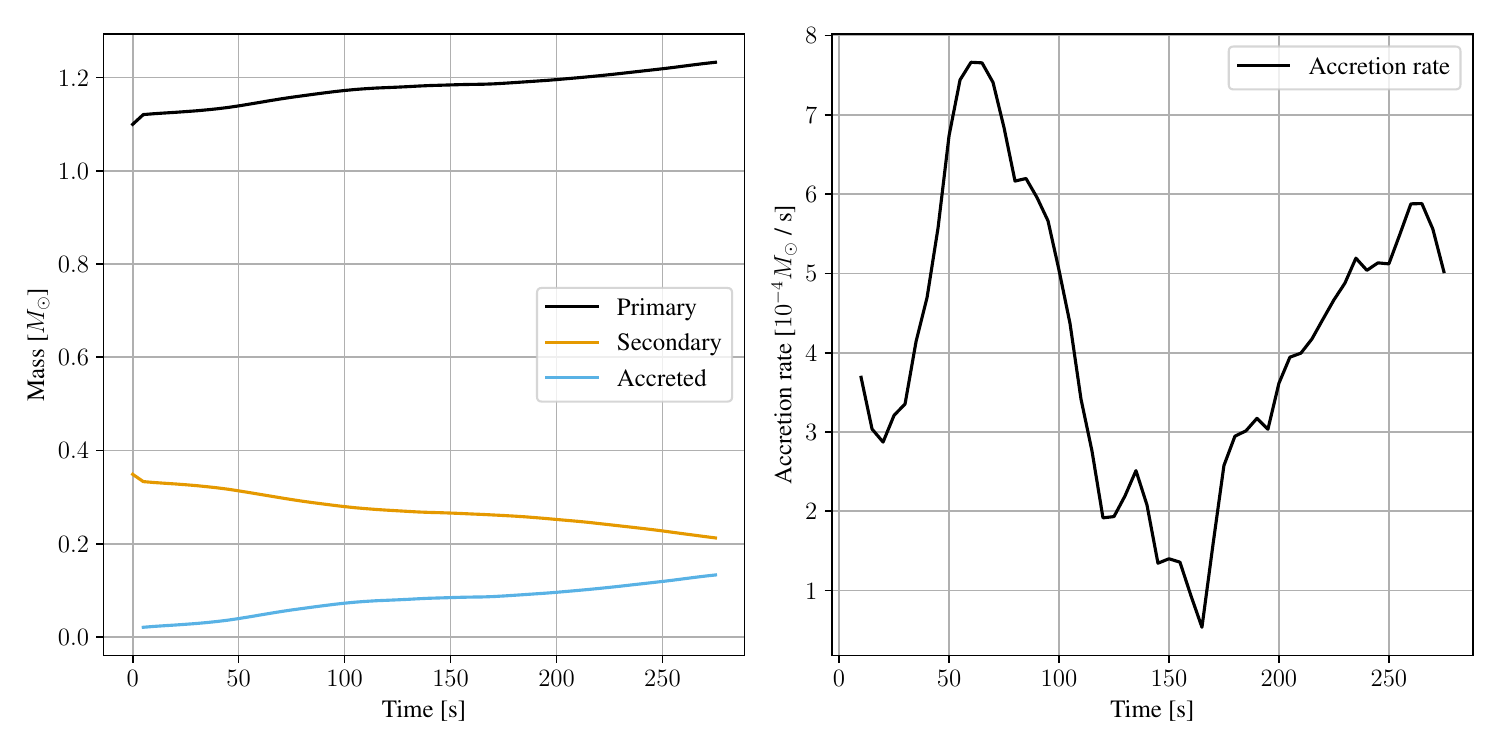}
    \caption{Accretion onto the primary as well as mass loss of the secondary is presented on the left, while the effective accretion rate is given on the right. Accreted particles are identified using the ``Passive Scalar'' labels which identify the particle's origin. Bound particles originating from the secondary which are now primarily under the gravitational influence of the primary are classified as accreted. The area of gravitational influence is delineated by the radius of the inner Lagrangian point. The points of maximum and minimum accretion on the right subplot correspond to the closest and furthest distances between the two stars in their orbit.}
    \label{fig:accretion_rate}
\end{figure*}

\begin{figure*}
    \centering
    \includegraphics[width=0.8\linewidth]{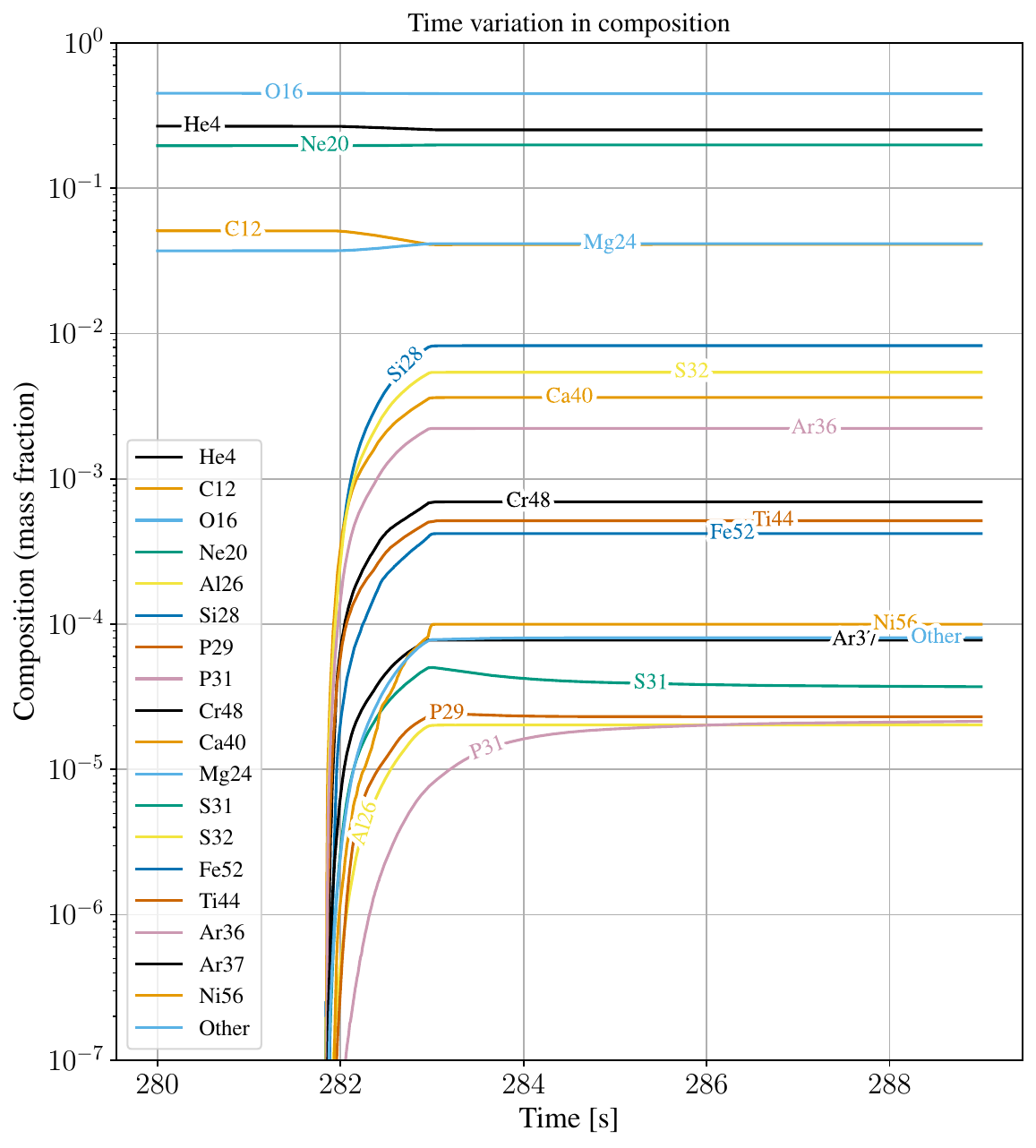}
    \caption{Time-variation of the composition of all simulated mass (primary \& secondary) for case C-120. The $y$-axis gives the percentage of mass made up by a specific species. ``Other'' indicates the summation of all species in the network not listed in the legend.}
    \label{fig:comp_evolution}
\end{figure*}

In this work, we have employed realistic chemical profiles for the primary WD by using a state of the art hydrogen-deficient WD model of 1.1\,\msol\ kindly supplied to us by Camisassa \citep[][C19 hereafter]{camisassaEvolutionUltramassiveWhite2019}. A diagram of the radial profile of C19 is shown in Figure~\ref{fig:maria_profile}. The reason for this choice is that nuclear reaction rates depend strongly on chemical composition. The initial chemical profiles of the C19 models correspond to progenitors whose initial Zero Age Main Sequence (ZAMS) masses are in the range $9-10.5$\,\msol. Their evolution to the final WD stage is as calculated by \citet{siessVizieROnlineData2010} and includes the semi-degenerate carbon burning that occurs during the pulsing Super AGB (SAGB) evolutionary phase. Hence, these models exhibit a self-consistent ONe profile and intershell masses built up during the SAGB. Importantly, for the present calculations, the detailed evolution through the SAGB phase yields realistic values of the WD's entire helium content. In our simulations, the C19 ONe\,WD has a thin He layer (of approximate mass $4 \times 10^{-4}$ \msol), a total mass of 1.1\,\msol\ and an effective temperature of approximately $3,300$\,\K. This 1D stellar structure was then ported into \texttt{AREPO} and transformed into a 3D structure with HEALPIX (Hierarchical Equal Area isoLatitude Pixelisation) mapping \citep{gorskiHEALPixFrameworkHighResolution2005}. The method of constructing shells with an appropriate width to reproduce the 1D profiles is described in further detail by \citet{pakmorStellarGADGETSmoothed2012}.

We have employed an isothermal, constant composition approximation for the pure-helium, low-mass (0.35\,\msol) secondary WD. We set the isothermal temperature to be $T = 5 \times 10^5$\,\K and generated its density and energy profiles using a simple numerical integrator \citep{pakmorFateSecondaryWhite2022}. This integrator estimates the central density using the total mass and then integrates the equations of hydrostatic equilibrium up to a cutoff density of $\rho_c = 1 \times 10^{-4}$\,\gcmcubed. Additionally, we also generate an isothermal, constant composition primary of mass 1.1\,\msol\ for comparison with the C19 models. We select an effective temperature of $T = 5 \times 10^5$\,K and use approximately the same overall homogeneous composition of 5\% $^{12}\mathrm{C}$, 55\% $^{16}\mathrm{O}$, 30\% $^{20}\mathrm{Ne}$, and 10\% $^{24}\mathrm{Mg}$.

A relaxation method was applied to stabilise the structures of the two WDs that may have suffered from discretisation errors and disruption of their hydrostatic equilibria when transformed from 1D to 3D structures \citep{ohlmannConstructingStable3D2017}. The WD structures were then evolved for approximately 80\,\s\, to confirm stability and check for energy conservation (see Paper\,I for further details). The two WDs were then set in a non-interacting co-rotating circular orbit with an initial orbital period of $T = 125$\,\s \, and a separation of $a = 4.26 \times 10^9$\,cm. The simulation box had a side length of $1 \times 10^{12}$\,cm. The density of the background mesh was set to $10^{-4}$\,\gcmcubed \, to prevent numerical issues arising from steep gradients from the stellar surface to empty space. ``Passive Scalars'' were used to track quantities such as the amount of accreted mass, by labelling all cells as initially belonging to one of the two WDs. 

Unfortunately, our simulations cannot feasibly follow the mass exchange from the onset of Roche-lobe overflow through a long period of gravitational radiation. This is due both to the amount of real-world time to follow the dynamics in full and the accumulation of numerical errors which would render the resulting simulation unreliable. Thus, we must resort to a method of approximating the result of this period of evolution. As described previously in Paper\,I and in \citet{pakmorThermonuclearExplosionMassive2021}, we can simulate the loss of angular momentum due to gravitational radiation at an accelerated rate. By adding a tidal force which decreases the separation of the stars at a constant rate, we can allow the stars in the binary to interact while passing through physically-reasonable states along the way. This allows both members of the binary to adapt to each other's potential as time goes on and it allows us to pick out moments when we can examine the dynamics in real time. This \texttt{INSPIRAL} routine is implemented in \texttt{AREPO} natively - the amount of angular momentum removed from the system is controlled by the user-set ``inspiral velocity'' and the amount of time for which the routine is run, the ``inspiral time''. We use a velocity of $50$\,\kmpersec which is active for about 1.5 orbits and decreases the separation from an initial value of $a = 4.26 \times 10^9$\,cm. The exact inspiral times used for each model are discussed in section~\ref{sec:results}. We note that \texttt{AREPO} does not contain the functionality to perform radiative transfer calculations during the real-time evolution phase for arbitrary input, so our simulations do not include this avenue of energy loss.

\section{Simulation Results} \label{sec:results}

\subsection{Isolated Phase and ONe WD structure stability} \label{sec:iso_phase}

We validate our transformation of the 1D spherically-symmetric model to 3D \texttt{AREPO} point structure by examining radial profiles of the 3D structure and comparing them to the originals. This process was repeated at various phases throughout the hydrodynamic evolution of the binary structure. These radial profiles confirmed that the basic structure was maintained, but with low-mass regions preserving the WD atmosphere less closely. In particular, \texttt{AREPO} cells in the atmosphere are larger and adjacent to the background grid, which is orders of magnitude lower in density than the WD atmosphere. This resulted in some leakage of mass between the exterior WD's layers and the low-density background ($\rho \leq 10^3$ \gcmcubed), decreasing the central density by approximately $6$\%. Because these high-mass ONe\,WD structures are much more compact than the CO\,WDs used in Paper\,I, we have also observed oscillatory changes, decreasing in magnitude over time, in the maximum radius as the structure settled.

\begin{table*}
\large
\centering
\caption{A summary of the models simulated, including the source of the 1D profile, and inspiral time. The network size for all simulations is 55-species. Here $a_2$ and $T_2$ are the orbital separation and period at the end of the inspiral phase, respectively. $E_{det}$ is the estimated explosion energy. The helium detonation, which is common to all models, begins $t_3$ seconds after the end of the inspiral phase. DT refers to the type of detonation. We discuss the detonation types further in the text.}
\label{tab:models_list}
\resizebox{\textwidth}{!}{
\begin{tabular}
{p{0.15\linewidth}p{0.10\linewidth}p{0.10\linewidth}p{0.15\linewidth}p{0.05\linewidth}p{0.05\linewidth}p{0.4\linewidth}}
\hline\\
1D Model Type & \(a_2\) [km] & \(T_2\) [s] & \(E_{\text{det}}\) [10\textsuperscript{49} erg] & \(t_3\) [s] & DT & Notes \\ 
\hline
C-120 & 35600 & 96.1 & 8.41 & 282 & .Ia & Both primary and secondary stars survive and accretion resumes after less than one orbit. \\
C-150 & 33400 & 87.6 & 5.77 & 170 & .Ia & Primary survives, secondary is effectively shredded by long inspiral \\
C-180 & 29200 & 71.4 & 6.08 & 140 & .Ia & Primary survives, secondary is effectively shredded by long inspiral \\
\hline
H-120 & 35600 & 96.3 & 5.27 & 196 & .Ia & Both primary and secondary stars survive and accretion resumed. Secondary very elongated. \\
H-150 & 33000 & 85.8 & 96.7 & 268 &  Ia & Both primary and secondary destroyed in an energetic supernova \\ 
\hline
\end{tabular}
}
\end{table*}

\subsection{Inspiral, Evolution and Merger} \label{sec:detonation}

We have run several simulations using different inspiral times (see Table~\ref{tab:models_list}). We decided on the inspiral times $T = 120$ and $T = 150$\,s because we had initially hoped to compare the outcomes of the current simulations with the $1.1$\,\msol\, CO simulations conducted in Paper I. Therefore these simulations used these same inspiral time values. In addition, we measured the L1 density and use this as a proxy for the correct time to terminate the inspiral routine - based on past experience, a value approximately in the range of $\rho_{L1} = 1 - 2$\,\gcmcubed\, is the first viable time to turn off the inspiral routine. We found a density value of $\rho_{L1} = 2.17$\,\gcmcubed\, for $T = 120$\,s, slightly above our target.

Firstly, we examine what the outcome of the merger is when inspiral is stopped after $T = 120$\,s leading to a separation $a = 3.56 \times 10^9\,\mathrm{cm}$ (with an effective period of $96.1$ \s). We denote this model with C-120 and depict its time evolution in Figure~\ref{fig:evolution}. The position coordinates denote the displacement from the centre of the simulated box. The first frame shows the state of the binary at the end of the inspiral phase $(t=0)$. Accretion begins slowly but accelerates as the secondary loses mass and the material escaping from the inner Lagrangian point forms an accretion disk around the primary WD (top two panels). 

Compare this with Figure~\ref{fig:accretion_rate}, which computes the amount of accreted mass based on the estimated location of the inner Lagrangian point and the masses of the two stars over time (left panel of Figure~\ref{fig:accretion_rate}).  We can also see that a average accretion rate is $4.54 \times 10^{-4} M_{\odot}/\mathrm{s}$, though this rate does vary throughout the orbit (right panel of Figure~\ref{fig:accretion_rate}). The instantaneous accretion rate is expected to be correct for the system, but the actual amount of accreted mass may be an underestimate due to the accelerated inspiral phase. That is, the true system would have been accreting material at a lower mass transfer rate but for a much longer period of time, which we cannot capture due to practical constraints, as discussed in section~\ref{sec:methods}.

Approximately $0.13$\,\msol\ is accreted before we observe the onset of nuclear burning at the base of the ONe primary's helium layer, at a radius of $r = 4.06 \times 10^{8}$\,\cm. By examining the cells around this point, we can determine the value of the primitive variables just prior to the helium detonation, though we cannot fully resolve this down to the length scale of centimetres. These values are: $\rho = 6.7 \times 10^5$\,\gcmcubed, $p = 2.0 \times 10^{22}$\,\barye, $T = 2.9 \times 10^{8}$\,\K. The approximate chemical composition in this neighbourhood is: $ X(^{12}\mathrm{C}) = 0.16, X(^{16}\mathrm{O}) = 0.20, X(^{4}\mathrm{He}) = 0.59$, $X(^{20}\mathrm{Ne}) = 0.04$, and $X(^{24}\mathrm{Mg}) = 0.01$. During this phase we use the 13-species nuclear network and a coarse snapshot output frequency owing to the computational and storage cost of the experiment. When a detonation was observed to begin, the simulation was restarted using the 55-species network and an output frequency of one snapshot every $0.25$\,\s at the last snapshot prior to helium ignition (i.e. $t = 280$\,\s). 

Viewing the third and fourth panel in Figure~\ref{fig:evolution}, we can see that the burning region envelops the high density core and intersects itself, thus causing an increase in the density and temperature on the opposite side of the primary. The burning region continues to propagate outwards, blowing away large portions of the accretion disk and shocking the companion star. During this episode, approximately $0.103$\,\msol\, is ejected from the binary. This resembles a sub-luminous supernova event because the burning is confined to the exterior layers of the WD whilst the ONe core remains intact. The quantitative connection with sub-luminous supernova events will be discussed further in section \ref{sec:discussion}. The shock converges at a radius $r = 2.1 \times 10^{8}$\,\cm, where the values of the primitive variables just before convergence are: $\rho = 1.5 \times 10^7$\,\gcmcubed, $p = 1.5 \times 10^{24}$\,\barye, $T = 2.8 \times 10^{8}$\,\K. The approximate chemical composition in this neighbourhood is: $ X(^{12}\mathrm{C}) = 0.05, X(^{16}\mathrm{O}) = 0.63, X(^{4}\mathrm{He}) = 0.00$, $X(^{20}\mathrm{Ne}) = 0.27$, and $X(^{24}\mathrm{Mg}) = 0.05$. Just after the shock convergence, the temperature and density are raised to $\rho = 4.5 \times 10^7$\,\gcmcubed, $T = 8.0 \times 10^{8}$\,\K.

\begin{figure*}
    \centering
    \includegraphics[width=\linewidth]{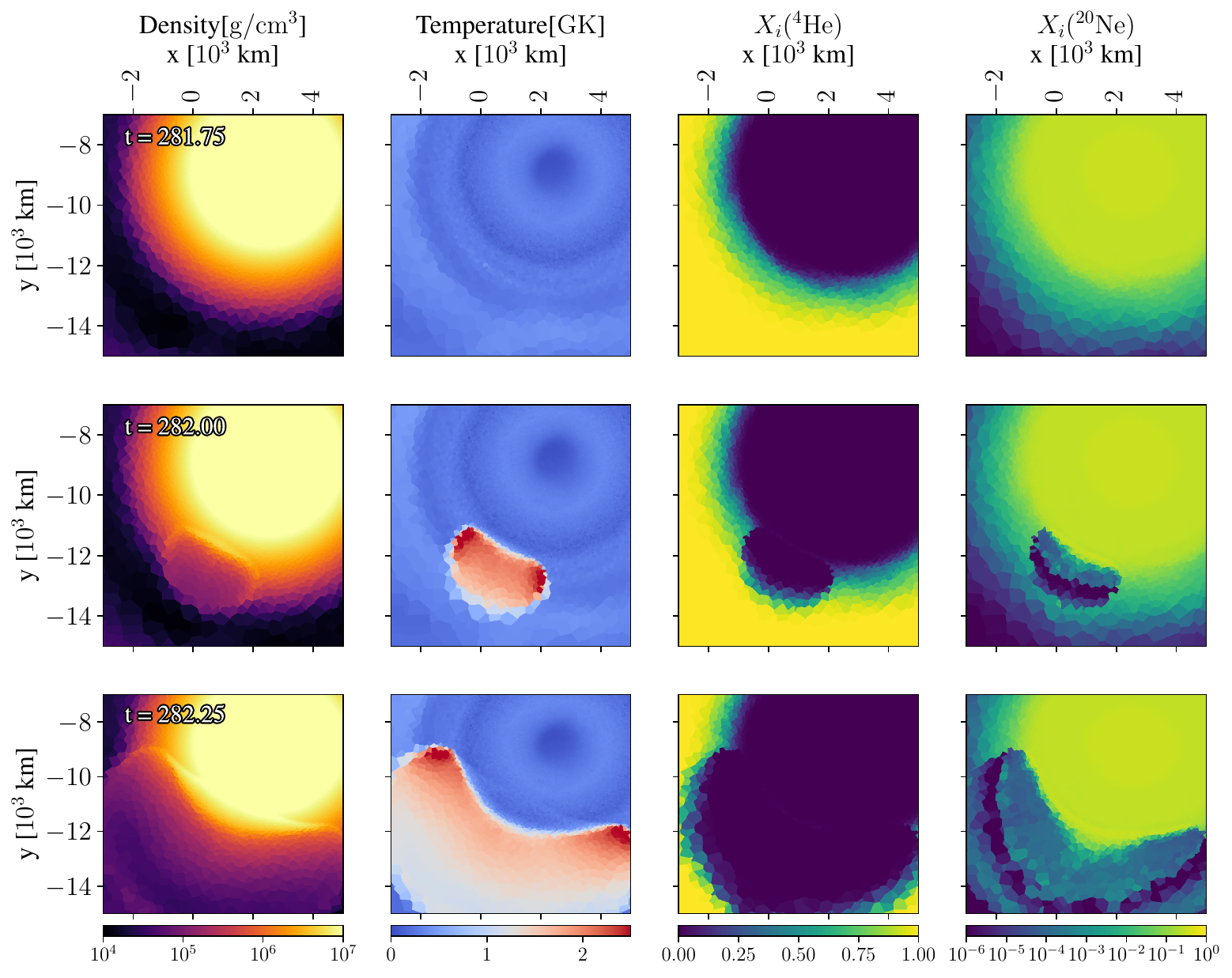}
    \caption{Location of primitive variables and species at the onset of the helium detonation at the base of the helium layer.}
    \label{fig:xnuc_pcolor}
\end{figure*}

Figure~\ref{fig:comp_evolution} shows the evolution of the mass fractions of all particles within the simulated box. The onset of burning begins near $t=282$ when $^{12}\mathrm{C}$ is converted to primarily intermediate mass elements, including $^{28}\mathrm{Si}$, $^{32}\mathrm{S}$ and $^{40}\mathrm{Ca}$. We have included a view of the initial moments before and after $t=282$\,\s\, in Figure~\ref{fig:xnuc_pcolor}, which shows the evolution of the burning region with high time resolution. The leading front of the burning region which proceeds around the primary has a maximum temperature of approximately $2.5\, \mathrm{GK}$ and primarily consumes $^{12}\mathrm{C}$ and $^{4}\mathrm{He}$ in the outer regions of the primary and its accretion disk.

By taking the difference of the nuclear composition of all simulated cells just before and after the explosion, we are able to estimate the energy release in the explosion, which we compute as $8.41 \times 10^{49}$ \,\erg. By comparing the centre of mass velocities of the primary just before and after detonation, we are able to compute the kick velocity provided by the detonation. We find these values to be $v_{b} = [212.0,  80.0, -0.19]$ \kmpersec, $v_{a} = [192.0, 135.0, -0.87]$\,\kmpersec. This gives us a kick velocity of $[-20.0, 55.0, 0.68]$\,\kmpersec, $v = 58.7$\,\kmpersec. These numbers can be compared to the detonation of the binary in Paper I - a 1.1\,\msol\, CO\,WD with a 0.35\,\msol\, secondary He\,WD. The outcome of that simulation was the complete destruction of the primary, releasing $1.61 \times 10^{51}$\,\erg with a kick velocity of $v = 301$\,\kmpersec. Obviously, due to the survival of the primary, this reaction is far less energetic. 

\begin{figure*}
    \centering
    \includegraphics[width=\linewidth]{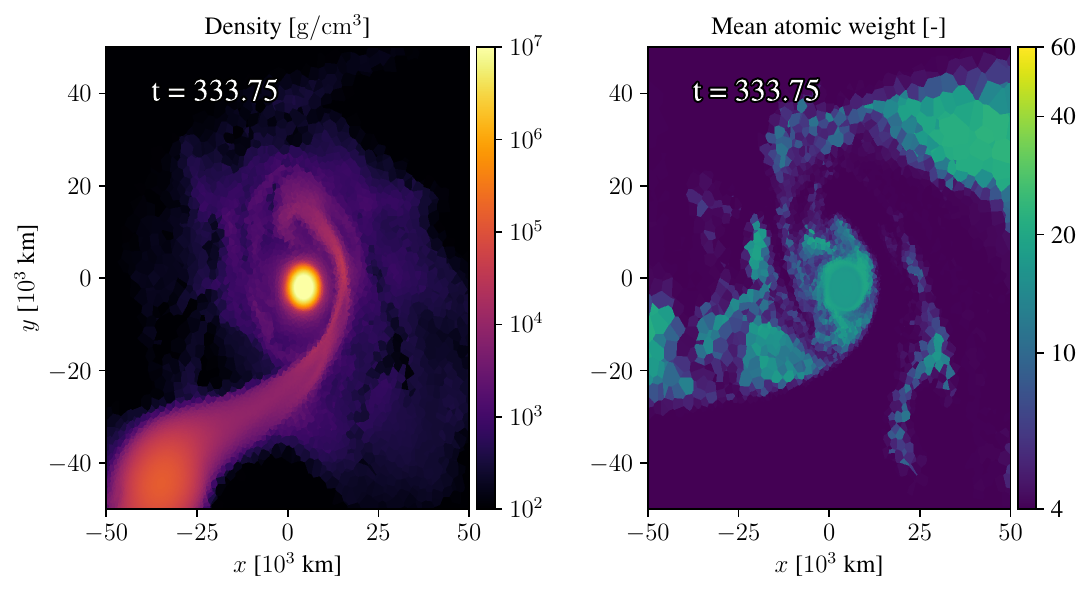}
    \caption{A map of the heavy ejecta and surviving binary for the C-120 case approximately 50 seconds after the detonation has encircled the primary. We observe that the secondary (at the bottom left of the left-hand side plot) has begun to transfer matter onto the primary again after being shocked. The location of the heavy ejecta is shown on the right-hand plot where the dark regions represent pure helium, while the lighter regions indicate areas enriched by heavier species. We also observe that some of these heavier species remain on the primary as contaminants.}
    \label{fig:pcolor_late_time}
\end{figure*}

The post-detonation state of the ejecta and binary are shown in Figure~\ref{fig:pcolor_late_time}, approximately 50\,\s after the detonation. We observe that the $0.103$\,\msol\, has been ejected into the surrounding region, including $7.09 \times 10^{-2}$\,\msol\, of $^{4}\mathrm{He}$ and various heavier elements. For this short inspiral scenario, the secondary remains intact and begins to accrete onto the primary again after less than one orbit. We have continued the simulation of the C-120 case to $t=650$ (i.e. approximately $370$ seconds beyond the original detonation time) using the 13-species network to observe whether any further detonations would take place which might indicate periodic behaviour in observations. No further detonations were observed until the end of our simulation, at which point the system had still not completely merged. 

Secondly, we examine what the outcome of the merger is when the inspiral phase is longer. In this experiment the inspiral is stopped after $T = 150$\,s leading to a separation $a = 3.34 \times 10^9\,\mathrm{cm}$ (with an effective period of $87.6$ \s). We denote this model with C-150. This case follows much the same trajectory for the majority of its evolution and also ends with a thermonuclear explosion which begins at the base of the helium layer and encircles the primary. Comparing the cases in Table~\ref{tab:models_list}, we can see that C-150 ends the inspiral phase with a smaller separation between the two stars and explodes more quickly (at $t=170$\s) than its C-120 counterpart ($t=282$\,s) due to the increased mass transfer rate. Because of the higher tidal forces experienced by the secondary under this scenario, a larger amount of mass is ejected in a tidal tail and with the secondary star almost completely shredded before the explosion takes place. The shock front of the expanding ejecta dissipates the secondary remnant further, resulting in a now isolated primary WD with its exterior layers highly contaminated by the nuclear burning ashes and surrounded by a very puffy accretion disk. 

We also investigated the response of the binary to a very long inspiral scenario, i.e. the inspiral is stopped after $T = 180$\,s\ leading to a separation $a = 2.92 \times 10^9\,\mathrm{cm}$ (with an effective period of $71.4$ \s). We denote this model with C-180. The outcome of this simulation was almost identical to the C-150 case. Due to the low initial distance and high mass transfer rate, the secondary is messily accreted onto the primary, resulting in a fluffy accretion disk and large tidal tail. We observe a shorter time to detonation ($t=140$\s), but the characteristics of the explosion are the same. The outer regions of the primary are affected and the secondary is significantly disrupted. 

As argued in Paper\,I, the inspiral time can be considered a parameter of the system which selects for the amount of angular momentum loss in the binary system so that the outcome of a merger can be simulated in a logistically-tractable amount of time. However, the cases with the shortest feasible inspiral times are in a sense the most realistic representations of two stars that begin to interact and merge under the influence of gravitational radiation. Thus, we will use the C-120 model as our canonical model for most figures and tables throughout the remainder of this paper. However, since the early interaction does not destroy the system, it is possible that the longer inspiral outcomes may be realised later in the binary's lifetime. Note also that all of the preceding simulations employ the chemical profiles from C19. We will also discuss the outcomes of the simulations employing the homogenous chemical structure later in section \ref{subsec:chemical_structure}.

\subsection{Ejecta} \label{sec:yield}

In Table~\ref{tab:bound_unbound_mass} we list the masses of the unbound ejecta surrounding the C19 models approximately $50$\,\s\, after detonation, as well as the bound mass which remains in the binary. This table was computed using the sum of the nuclear composition of each \texttt{AREPO} cells multiplied by the mass of that point, excluding box particles. We can see that the largest mass contribution in the ejecta consists of unburned $^{4}\mathrm{He}$, some of which was tidally ejected before the onset of burning. The next largest mass contributions are $^{16}\mathrm{O}$, $^{28}\mathrm{Si}$, $^{32}\mathrm{S}$ and $^{40}\mathrm{Ca}$. The $^{16}\mathrm{O}$ consists of a mixture of unburned fuel from the outer layers of the primary, and generated from the reaction $^{12}\mathrm{C}+^{4}\mathrm{He}\rightarrow ^{16}\mathrm{O}$. We can see in Figure~\ref{fig:comp_evolution} that the most substantial decrease in mass of the initial species is the drop in $^{12}\mathrm{C}$ and $^{4}\mathrm{He}$. The $1.26 \times 10^{-4}$\,\msol\, of ejected $^{56}\mathrm{Ni}$ is one of the most interesting potential observables due to its decay chain to $^{56}\mathrm{Fe}$, though the strong representation of Calcium and the other intermediate-mass elements is likely to be another interesting spectral feature. 

\begin{figure*}
    \centering
    \includegraphics[width=0.75\linewidth]{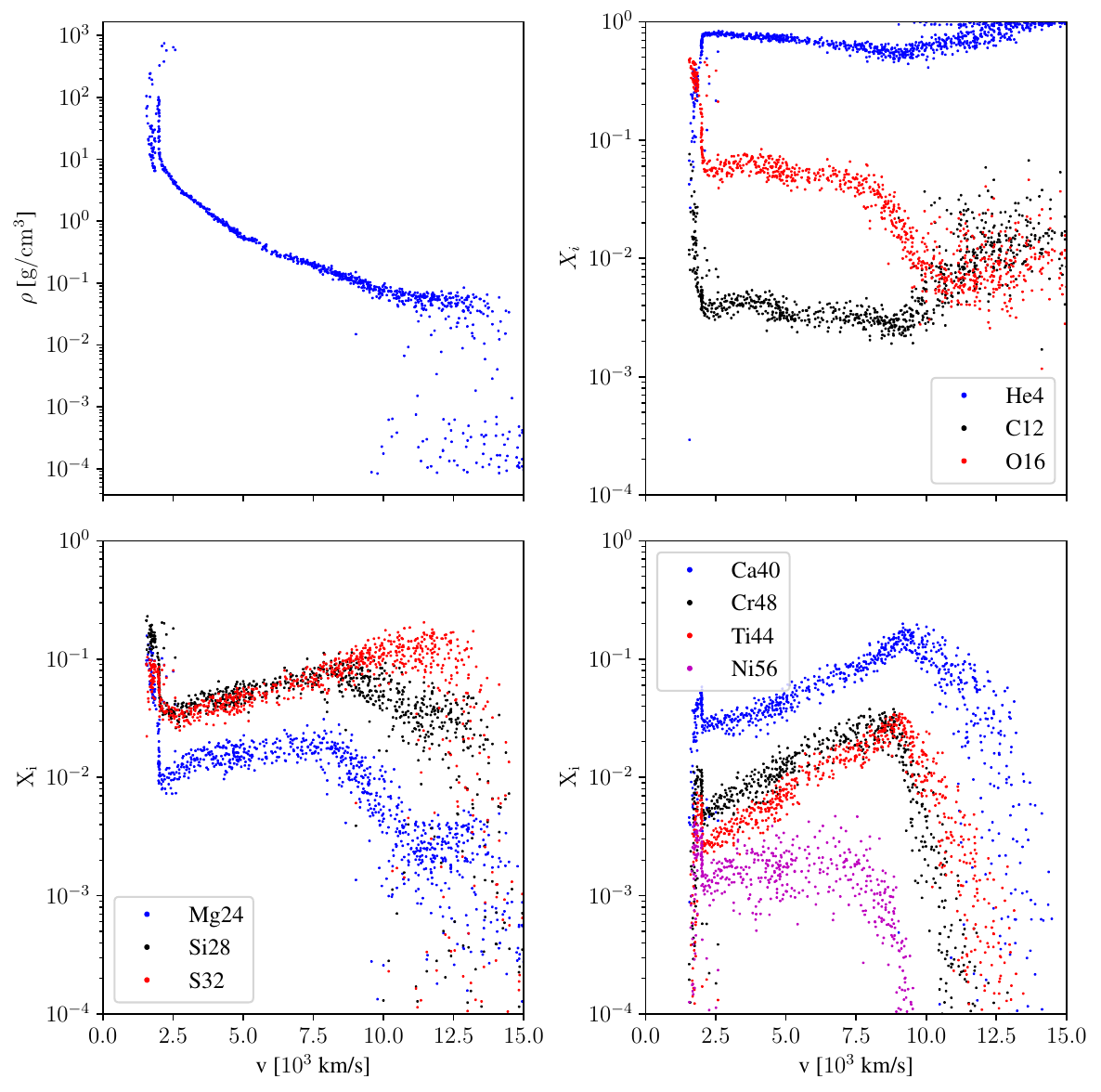}
    \caption{Distribution of densities and compositions in velocity space $\approx 50$\,\s \, after detonation for the C-120 model. The high-velocity region is dominated by unburnt $^4$He. The higher velocities are also dominated by lower density particles.}
    \label{fig:velocity_space}
\end{figure*}

We can see the distribution of elements as a function of velocity in Figure~\ref{fig:velocity_space}. The relatively low velocity range compared to SNe\,Ia is noteworthy, but not surprising given the low explosion energy. This is in contrast to the high-velocity ejecta in Paper I which reached velocities of $>30 \times 10^3$\,\kmpersec. We can also observe very high variability in the composition at low velocities - this is likely due to interactions with other objects such as the surviving secondary or accretion disk (see, for example, Figure~\ref{fig:pcolor_late_time}). As expected, high velocities are dominated by unburned helium, carbon and oxygen, as opposed to the heavier elements. We will compare C-120's yield and velocities to existing models in the following section.

\begin{table*}
\centering
\caption{Mass composition of bound and unbound particles compared between the various models approximately $50$\,\s\, after detonation, presented for selected species. Note that marginal values (e.g. the C-150 $^{56} \mathrm{Ni}$ values) should not be considered as reliable estimates and are included for completeness.}
    \label{tab:bound_unbound_mass}
    \resizebox{\linewidth}{!}{
    \begin{filecontents*}{data/table_boundedness_cutoff.txt}
Species,C120B,C120UB,C150B,C150UB,C180B,C180UB
He4,2.58e-01,7.09e-02,2.72e-01,6.18e-02,2.71e-01,6.20e-02
C12,6.12e-02,4.75e-04,6.73e-02,3.51e-04,6.58e-02,3.17e-04
O16,6.65e-01,6.49e-03,6.68e-01,1.76e-03,6.69e-01,2.94e-03
Ne20,2.97e-01,6.90e-04,3.01e-01,4.70e-04,3.01e-01,4.01e-04
Mg24,6.08e-02,1.67e-03,6.17e-02,1.21e-03,6.07e-02,9.96e-04
Si28,6.50e-03,5.92e-03,4.22e-03,4.50e-03,4.06e-03,4.09e-03
S32,2.24e-03,5.91e-03,5.47e-04,4.67e-03,7.85e-04,4.43e-03
Ar36,5.30e-04,2.81e-03,6.83e-05,2.40e-03,1.14e-04,2.17e-03
Ca40,3.96e-04,5.06e-03,1.31e-05,2.59e-03,7.67e-05,4.58e-03
Ni56,2.37e-05,1.26e-04,2.65e-14,9.71e-12,1.06e-08,1.82e-08
Other,3.14e-04,2.53e-03,7.17e-05,1.70e-04,1.06e-04,5.78e-04
Total,1.35e+00,1.03e-01,1.38e+00,7.99e-02,1.37e+00,8.25e-02
\end{filecontents*}
\pgfplotstabletypeset[
    multicolumn names,
    col sep=comma, 
    columns=
    {Species,C120B,C120UB,C150B,C150UB,C180B,C180UB},
    columns/Species/.style={string type, column name=Species}, 
    columns/C120B/.style={string type, column name=C-120 Bound [\msol], column type={|c}}, 
    columns/C120UB/.style={string type, column name=C-120 Unbound [\msol], column type={c|}}, 
    columns/C150B/.style={string type, column name=C-150 Bound [\msol], column type={c}}, 
    columns/C150UB/.style={string type, column name=C-150 Unbound [\msol], column type={c|}}, 
    columns/C180B/.style={string type, column name=C-180 Bound [\msol], column type={c}}, 
    columns/C180UB/.style={string type, column name=C-180 Unbound [\msol], column type={c}}, 
    every head row/.style={before row=\toprule, after row=\midrule}, 
    every last row/.style={after row=\bottomrule},
    ]{data/table_boundedness_cutoff.txt}
    }
\end{table*}

\section{Discussion} \label{sec:discussion}

\subsection{Sub-luminous supernovae from helium shell detonations}

Because our simulations did not result in the disruption of the ONe\,WD primary, they cannot be interpreted as a typical SNe\,Ia. Instead, they seem to describe a sub-luminous event with a much lower total ejecta mass than the typical SNe\,Ia. Such an event is unlikely to be a 1991bg-like SN\,Ia \citep{graurUnderluminous1991bglikeType2024, crockerDiffuseGalacticAntimatter2017}, as helium is not typically observed in these sub-luminous SNe, which are primarily thought to involve low-mass CO\,WDs, possibly with a thin helium shell, undergoing thermonuclear detonation.

A link between the outcomes of our simulations and the attributes of sub-luminous events is possible. However, we conclude SN\,Iax events are unlikely to fit the characteristics produced by our simulations. SNe\,Iax are a subclass of thermonuclear SNe thought to arise from deflagrations that fail to fully disrupt the WD \citep[see][for a review on SNe\,Iax]{jhaTypeIaxSupernovae2017}. They are characterised by low luminosities, slow expansion velocities ($\sim 4,000-10,000 \, \mathrm{km \, s^{-1}}$), incomplete C and O burning, and spectra displaying intermediate-mass elements (e.g., silicon, sulphur, calcium) alongside unburned carbon and oxygen. Typically, they eject $0.2-0.6$\,\msol\ of material, including $0.003-0.3$\,\msol\ of $^{56}$Ni. Furthermore, two SNe\,Iax, SN\,2004cs and SN\,2007J, exhibit helium lines in their spectra \citep{foleyTYPEIaxSUPERNOVAE2013}. 

Alternatively, we can compare to theoretical models of helium detonations such as the sub-luminous ``dot Ia'' (.Ia SNe) described in \citet{bildstenFaintThermonuclearSupernovae2007} and later \citet{shenUnstableHeliumShell2009}. Previous simulations predict these should be observed as faint, rapidly evolving thermonuclear transients resulting from the detonation of a helium shell on a WD. They are characterised by low luminosities, fast-evolving light curves, and the ejection of $<0.1$\,\msol\ of material, including $<0.01-0.05$\, \msol\ of $^{56}\mathrm{Ni}$. Our results should therefore be compared with those of \citet{shenThermonuclearIaSupernovae2010}, who explored the hydrodynamic and nucleosynthesis evolution of 1D spherically symmetric helium detonations on WDs accreting helium from a companion star. Unlike SNe\,Ia, which produce significant amounts of $^{56}\mathrm{Ni}$, \citet{shenThermonuclearIaSupernovae2010} showed that .Ia SNe predominantly produce heavy alpha-chain elements (from $^{40}\mathrm{Ca}$ to $^{56}\mathrm{Ni}$) and leave behind unburned helium. This influences their spectral signatures, which lack the intermediate-mass elements typical of SNe\,Ia and are instead dominated by features of Ca\,II and Ti\,II.

Possible progenitors of .Ia SNe include binaries resembling AM\,CVn systems, which consist of very low-mass WDs in tight orbital configurations with massive WD companions. \citet{kilicFoundProgenitorsAM2014} identified systems such as SDSS J075141.18-014120.9, containing a 0.19\,\msol\ WD and a 0.97\,\msol\ WD companion in a 1.9-hour orbit, and SDSS J174140.49+652638.7, containing a 0.17\,\msol\ WD with an unseen companion of at least 1.11\,\msol\ in a 1.5-hour orbit. These systems could evolve into interacting AM\,CVn systems potentially leading to thermonuclear SNe\,.Ia within $\sim 100$ million years. Observations and analysis of Supernova 2010X (SN\,2010X) also suggest it as a possible candidate for a .Ia explosion \citep{kasliwalRapidlyDecayingSupernova2010}. The spectral characteristics of SN\,2010X include high velocities (approximately 10,000\,\kmpersec) and a chemical composition indicating the presence of elements such as oxygen, calcium, carbon, and possibly helium or aluminium.

\citet{shenThermonuclearIaSupernovae2010} also provides us with a useful point of comparison in terms of simulated ejecta and velocities. This study employed simulations of isothermal ($T = 1 \times 10^7 K$) constant-composition cores composed of 50\% $^{12}\mathrm{C}$ and 50\% $^{16}\mathrm{O}$ surrounded by an isentropic envolope composed of pure helium. \citet{shenThermonuclearIaSupernovae2010} examined various combinations of core and shell mass, but the most similar to our case is $(M_{WD}, M_{env}) = (1.0, 0.1)$\,\msol. These simulations require an artificial ignition, created by perturbing the temperature at the base of the helium envelope by approximately 10\% at the beginning of the hydrodynamic simulation. This results in a thermonuclear runaway which steepens into a shock which produces $0.090$\,\msol\, of ejecta (for our chosen model).

The $(M_{WD}, M_{env}) = (1.0, 0.1)$\,\msol\, simulation results in an ejecta velocity of $1.2 \times 10^{4}$\,\kmpersec\, consistent with the higher-velocity ejecta shown in Figure~\ref{fig:velocity_space}. \cite{shenThermonuclearIaSupernovae2010} also provides a table of ejecta which is useful for comparison. They predict the following in solar masses: $m_{He4} = 0.0351$, $m_{Ar36} = 5.22 \times 10^{-6}$, $m_{Ca40} = 1.26 \times 10^{-4}$, $m_{Ti44} = 5.94 \times 10^{-4}$, $m_{Cr48} = 9.9 \times 10^{-4}$, $m_{Fe52} = 2.88 \times 10^{-3}$, and $m_{Ni56} = 0.0504$. Our simulations produce significantly less $^{56}\mathrm{Ni}$. Our simulations also produce roughly twice as much $^{4}\mathrm{He}$, but this is also likely due to differences in the initial structure - whereas \citet{shenThermonuclearIaSupernovae2010} use a structure which initially contains $0.1$\,\msol\, of helium, our structure accretes  $0.13$\,\msol\, of helium before detonation and also contains the remaining secondary which could be stripped of matter which will become ejecta. The other intermediate-mass elements are of similar orders of magnitude. Regarding the feasibility of the mass transfer rate, \citet{bildstenFaintThermonuclearSupernovae2007} give an upper estimate for the mass accretion rate of such systems as $1 \times 10^{-7} M_{\odot}/\mathrm{yr}$. If mass transfer occurred at this rate for tens of millions of years, the secondary would already be substantially reduced by the time of the helium detonation. 

Our results also bear some similarity with calcium-rich transients observed in the past decades. For example, \citet{jacobson-galanCaHnkCalciumrich2020} report on SN 2016hnk, a Ca-rich supernova which appears consistent with models of a He-shell detonation on a CO WD. \citet{jacobson-galanCaHnkCalciumrich2020} model the spectra and light-curve of this SN and determine the most likely progenitor to be the detonation of a $0.02$\, \msol\, He-shell covering a $0.85$\, \msol\, CO WD. This transient had a relatively low luminosity and produced $0.03 \pm 0.01$\, \msol\, of $^{56}\mathrm{Ni}$ and a total ejecta of $0.9 \pm 0.3$\, \msol. Again, this is significantly more nickel than has been produced by any of our models, but the estimated total ejecta is well within the estimated range. \citet{deZwickyTransientFacility2020}, meanwhile, report on efforts to create a volume-limited sample of Calcium-rich Gap Transients using the Zwicky Transient Facility (ZTF) alert stream. They conclude that Ca-Ia and red Ca-Ib/c events arise from different progenitor systems than the green Ca-Ib/c events. The former are attributed to the double detonation of Helium shells and the latter with low-efficiency burning such as would result in a deflagration. Notably, \citet{deZwickyTransientFacility2020} conclude the ZTF events are most consistent with explosive burning on low-mass WDs, while our ONe primary is towards the upper end of the mass range. 

The major distinction to be drawn between our simulations, the observed transients and the .Ia SNe of \citet{shenThermonuclearIaSupernovae2010} is the total nickel production. At best, our models in Table~\ref{tab:bound_unbound_mass} currently produce $\sim 10^{-4}$\,\msol\ of $^{56}\mathrm{Ni}$, roughly two orders of magnitude smaller than the rest of our comparison class. At the least, we would expect our supernovae to be far less luminous. In summary, the characteristics of the less energetic detonation observed in our simulations, combined with the putative progenitor system of .Ia SNe (an AM\,CVn system whose stellar masses and separation are in general agreement with our pre-explosion binary's parameters) suggest that our simulations will resemble .Ia SN, albeit far less luminous.

\begin{figure*}
    \centering
    \includegraphics[width=\linewidth]{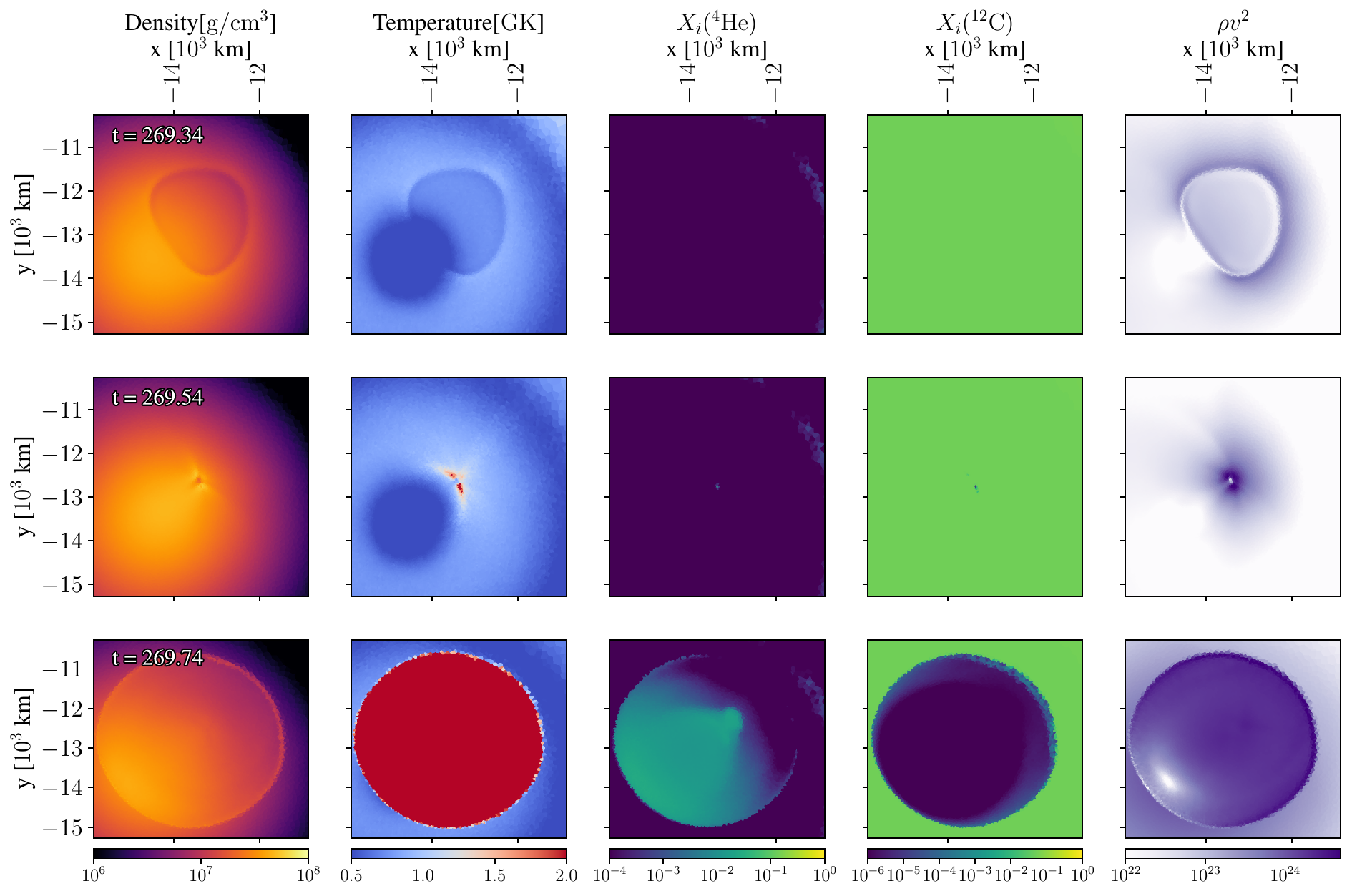}
    \caption{Density, temperature, and chemical composition against time in the $xy$-plane (the orbital plane of the binary) for the simulations of a chemically homogeneous structure with inspiral time of $150$\,s. The quantity $\rho v^2$ is also included -- this is a proxy for the kinetic energy of the cells and is useful in locating the convergence point and shock front of a detonation. This shock convergence in the interior is characteristic of the scissors mechanism of detonation (see section \ref{subsec:chemical_structure}).}
    \label{fig:h150_convergence}
\end{figure*}

\begin{figure*}
    \centering
    \includegraphics[width=\linewidth]{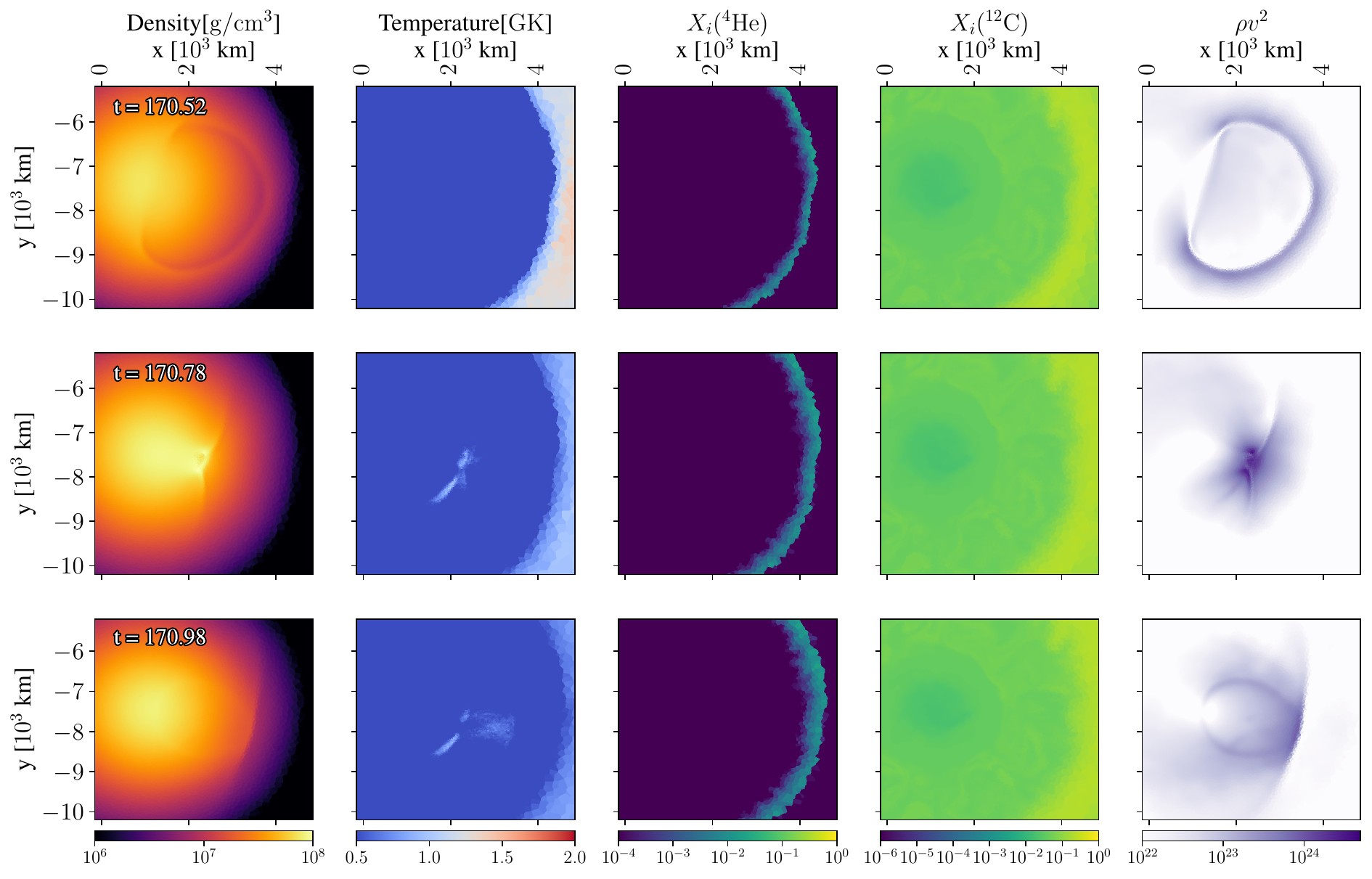}
    \caption{Density, temperature, and chemical composition against time in the $xy$-plane (the orbital plane of the binary) for the simulations of the C-150 model. See the previous figure for further explanation of the $\rho v^2$ quantity.}
    \label{fig:c150_convergence}
\end{figure*}

\subsection{Dependence of results on the chemical structure of the WD and on the inspiral time} \label{subsec:chemical_structure}

We now explore the dependence of our results on changes in the composition profile of the WD by comparing simulations using structures obtained by using the isothermal, homogeneous composition profile of a $1.1$\,\msol\ to a model of the same mass obtained using a realistic WD's profile computed by C19 (see section \ref{sec:models}). The simulations were performed using the inspiral parameters provided in Table~\ref{tab:models_list}. Firstly, we examine the results obtained when using the 55-species nuclear network and when inspiral is stopped after $T = 120$\,s (model H-120) and after $T = 150$\,s (model H-150). The same $0.35$\,\msol\, secondary (as described in section \ref{sec:models}) is used for both these cases. We have found that H-120 resembles closely enough the outcome of C-120 discussed in section \ref{sec:models} obtained with the WD structure of C19. Once again, there is a helium detonation which is ignited at the base of the helium layer and burns the outer regions of the primary. After the ejection of the burned material, accretion begins again, but the whole binary has been significantly altered by the previous interaction. 

By contrast, the H-150 simulation has shown significantly different behaviour, as evidenced by its significantly higher explosion energy in Table~\ref{tab:models_list}. As with the previous simulations, a helium detonation is triggered at the base of the helium layer ($r = 5.2 \times 10^{8}$\,\cm). These values of the primitive variables in this neighbourhood are: $\rho = 3.0 \times 10^5$\,\gcmcubed, $p = 1.0 \times 10^{22}$\,\barye, $T = 5.0 \times 10^{8}$\,\K. The approximate chemical composition in this neighbourhood is: $ X(^{12}\mathrm{C}) = 0.08, X(^{16}\mathrm{O}) = 0.30, X(^{4}\mathrm{He}) = 0.39$, $X(^{20}\mathrm{Ne}) = 0.17$, and $X(^{24}\mathrm{Mg}) = 0.06$. Again we observe that the shockwave travels through and around the primary. 

However, in this simulation, a secondary detonation is triggered at the point of shock convergence ($r = 1.37 \times 10^{8}$\,\cm). This is visible in Figure~\ref{fig:h150_convergence}, and appears to follow the ``scissors mechanism'', described by \citet{gronowSNeIaDouble2020}. These average values of the primitive variables in this neighbourhood are before the shock convergence are: $\rho = 2.0 \times 10^7$\,\gcmcubed, $p = 2.4 \times 10^{24}$\,\barye, $T = 7.1 \times 10^{8}$\,\K. The approximate chemical composition in this neighbourhood is: $ X(^{12}\mathrm{C}) = 0.05, X(^{16}\mathrm{O}) = 0.55, X(^{4}\mathrm{He}) = 0.00$, $X(^{20}\mathrm{Ne}) = 0.30$, and $X(^{24}\mathrm{Mg}) = 0.10$.  Just after the shock convergence, the temperature and density are $\rho = 4.7 \times 10^7$\,\gcmcubed, $T = 1.4 \times 10^{9}$\,\K, as a high temperature region begins to consume the primary. 

This mechanism is thought to be one of the key pathways leading to double detonations of WD binaries. \citet{gronowSNeIaDouble2020} describes how a helium detonation ignites within the accreted helium shell surrounding the WD. The detonation wave travels across the star’s surface and converges at a point opposite the initial ignition site. This convergence generates a shock that compresses the underlying CO core, potentially igniting a secondary carbon detonation. The term ``scissors mechanism'' refers to the way the detonation waves move and collide, resembling the closing motion of scissor blades. The X-shaped scissors are visible in our simulations that use a homogeneous and isothermal WD structure (see section \ref{sec:models}), an inspiral phase of 150\,s (see the second row Figure~\ref{fig:h150_convergence}). Note also that we can compare the characteristics of the shock convergence with the C-150 case which is shown in Figure~\ref{fig:c150_convergence}, as it has the same inspiral time. The shock convergence occurs at a very similar radius ($r = 1.32 \times 10^{8}$\,\cm) and the primitive variables are as follows: $\rho = 3.4 \times 10^7$\,\gcmcubed, $p = 4.9 \times 10^{24}$\,\barye, $T = 2.7 \times 10^{8}$\,\K. The approximate chemical composition in this neighbourhood is: $ X(^{12}\mathrm{C}) = 0.035, X(^{16}\mathrm{O}) = 0.61, X(^{4}\mathrm{He}) = 0.00$, $X(^{20}\mathrm{Ne}) = 0.306$, and $X(^{24}\mathrm{Mg}) = 0.05$. As is visible in Figure~\ref{fig:c150_convergence}, the C-150 model does not experience a secondary detonation.

The success of this mechanism strongly depends on factors such as the composition and structure of the transition region between the He shell and the CO core (see paper\,I) or the He and the C shells in the present calculations (see the chemical abundances as a function of radius in Figure~\ref{fig:maria_profile} derived from a realistic model for the internal structure of the WD). Mixing between the core and/or shell(s) during the accretion phase and/or while He-burning encircles the WD can significantly affect the likelihood of a secondary detonation travelling inwards and leading to a SN\,Ia event \citep[see][for more details]{gronowSNeIaDouble2020}. We believe the reason that this homogenous structure is capable of detonating -- in comparison to the models using the C19 WD structures -- is primarily due to the higher convergence temperature ($T = 1.4 \times 10^{9}$\,\K\, vs \, $T = 8.0 \times 10^{8}$\,\K) and higher number density of $^{12}\mathrm{C}$ ($0.05$ vs $0.035$). The carbon number density is particularly relevant due to the low temperature and density required for burning in comparison to oxygen. The higher temperature seems to be enough to overcome the fact that there is a slightly higher density at the C-150 convergence point. The higher temperatures at the convergence point may be due to a stronger helium detonation occurring in the homogenous case; this is supported by the fact that we observe more burning of the primary's initial species in the homogenous case, leading to a stronger shock compression. Nevertheless, since we cannot fully resolve the detonation down to the length scale of centimetres, we consider the secondary ignition to be a marginal event. That is, numerics may play a role under certain circumstances. 

\section{Conclusions} \label{sec:conclusions}

We have simulated WD binary mergers using realistic $1.1$\,\msol\, ONe, H-deficient chemical profiles for the primary WD computed by C19. These profiles were created using the stellar evolution code \texttt{LPCODE} and translated to a 3D \texttt{AREPO} structure using a custom \texttt{HEALPIX} mapping. After applying a relaxation process to mitigate discretisation errors, the 1.1\,\msol\, ONe WD was paired with a 0.35\msol\, isothermal, constant-composition He WD companion. To allow us to observe the interaction in a feasible amount of time, we employed the \texttt{INSPIRAL} routine which drains angular momentum from the binary similar to enhanced gravitational radiation, but at a much accelerated rate, for about 1.5 orbits. We simulated the interaction of these binaries after the accelerated inspiral using a hydrodynamic simulation in \texttt{AREPO}, coupled with a 13-species nuclear network. 

We delineated our simulations by the length of the inspiral time, creating the C-120, C-150 and C-180 simulations using the same realistic profile from C19. We have argued that the slowest accretion simulations are the most realistic, so our C-120 model is considered the canonical outcome, though the longer-inspiral cases may also occur at a later time during the binary's lifetime. Our simulations have shown that a thermonuclear runaway explosion is triggered at the base of the helium layer enveloping the ONe\,WD primary and converging on the opposite side. The binary survives in the C-120 case and begins accreting again after less than one orbit, though no further thermonuclear runaways are observed. The secondary is disrupted by strong tidal forces in the C-150 and C-180 cases, resulting in a surviving WD with a large accretion disk. 

Our canonical C-120 simulation produces $0.103$\,\msol\, of ejecta with a maximum velocity of approximately $15.0 \times 10^3$\,\kmpersec, with a strong drop-off in heavier elements at $12.5 \times 10^3$\,\kmpersec. The ejecta thrown into space are dominated by $^{4}\mathrm{He}$, $^{16}\mathrm{O}$, $^{28}\mathrm{Si}$, $^{32}\mathrm{S}$, and $^{40}\mathrm{Ca}$, with an unbound $^{56}\mathrm{Ni}$ mass of $1.26 \times 10^{-4}$\,\msol. We compared our results to several different simulation results and observations, including SNe\,Iax \citep{jhaTypeIaxSupernovae2017} and SNe\,.Ia \citep{shenThermonuclearIaSupernovae2010}. By comparing the ejecta mass, composition and velocities, we conclude the progenitor system is similar to an AM\,CVn system detonating via the .Ia pathway, notwithstanding orders of magnitude lower production of $^{56}\mathrm{Ni}$. In particular, we compare to the $(M_{WD}, M_{env}) = (1.0, 0.1)$\,\msol\, model of \citet{shenThermonuclearIaSupernovae2010}.

Lastly, we compared our results obtained using the chemical profiles of C19 to simulations using WDs with 1D constant-composition profiles with the same mass created using a numerical integrator. The H-120 simulation showed the same detonation pathway and a similar energy to the C19 models. However, the H-150 model resulted in a double detonation via the x-scissor mechanism, with the density and temperature at the convergence point reaching $\rho = 4.7 \times 10^7$\,\gcmcubed, $T = 1.4 \times 10^{9}$\,\K\, after convergence. We attribute this detonation both to the very high temperature at the convergence point but also the higher number density of $^{12}\mathrm{C}$ when compared to the C-150 model ($0.05$ vs $0.035$) We consider this to be a cautionary result of the possible outcome of simulation with overly-simplistic chemical profiles and long inspiral times. Overall, we consider the consensus outcome of a .Ia detonation to be the most realistic. 

\section*{Code Availability}

The \texttt{AREPO} code is publicly available \href{https://arepo-code.org}{here}, however some features are only available on the private ``Development'' branch. Access to the development branch is subject to approval from \texttt{AREPO}'s development team. 

\section*{Software}

This work made use of the following open-source software projects: NumPy \citep{harrisArrayProgrammingNumPy2020}, MatPlotLib \citep{hunterMatplotlib2DGraphics2007}, SciPy \citep{virtanenSciPy10Fundamental2020}.

\section*{Acknowledgements}

The authors would like to especially thank Maria Camisassa for providing the ONe model which forms the basis of this work. This research was undertaken with the assistance of resources and services from the National Computational Infrastructure (NCI), which is supported by the Australian Government. Computation time was contributed by the Australian National University through the Merit Allocation Scheme (ANUMAS), and the National Computational Merit Allocation Scheme (NCMAS). U.P.B was supported by the Australian Government Research Training Program (AGRTP) Fee Offset Scholarship and ANU PhD Scholarship (Domestic). This work was supported in part by the Alexander von Humboldt Foundation via the Friedrich Wilhelm Bessel Research Award to I.R.S., which enabled the research stay at HITS during which key components of this study were developed.

\section*{Data Availability}

The data underlying this article will be shared on reasonable request to the corresponding author.


\bibliographystyle{mnras}
\bibliography{main}



\appendix

\section{Influence of the nuclear network} \label{app:influence_network}

Nuclear networks must be integrated into hydrodynamical simulations to identify the region where nuclear reactions first occur and whether they could trigger SN explosions leading to the total or partial disruption of the WD. Furthermore, nuclear networks allow us to track changes in nuclear species and the rate of energy production from nuclear reactions. This integration creates a feedback loop between the hydrodynamical conditions and nuclear reactions. However, because the nuclear reaction networks are expensive to evolve owing to their very stiff system of equations that require implicit solvers, the nuclear networks used in many simulations are often limited to a small number of chemical species (particularly in two or three dimensional simulations) to save computational time which may lead to inaccurate results \citep{garcia-senzNotForgetElectrons2024}. 

\begin{figure*}
    \centering
    \includegraphics[width=0.95\linewidth]{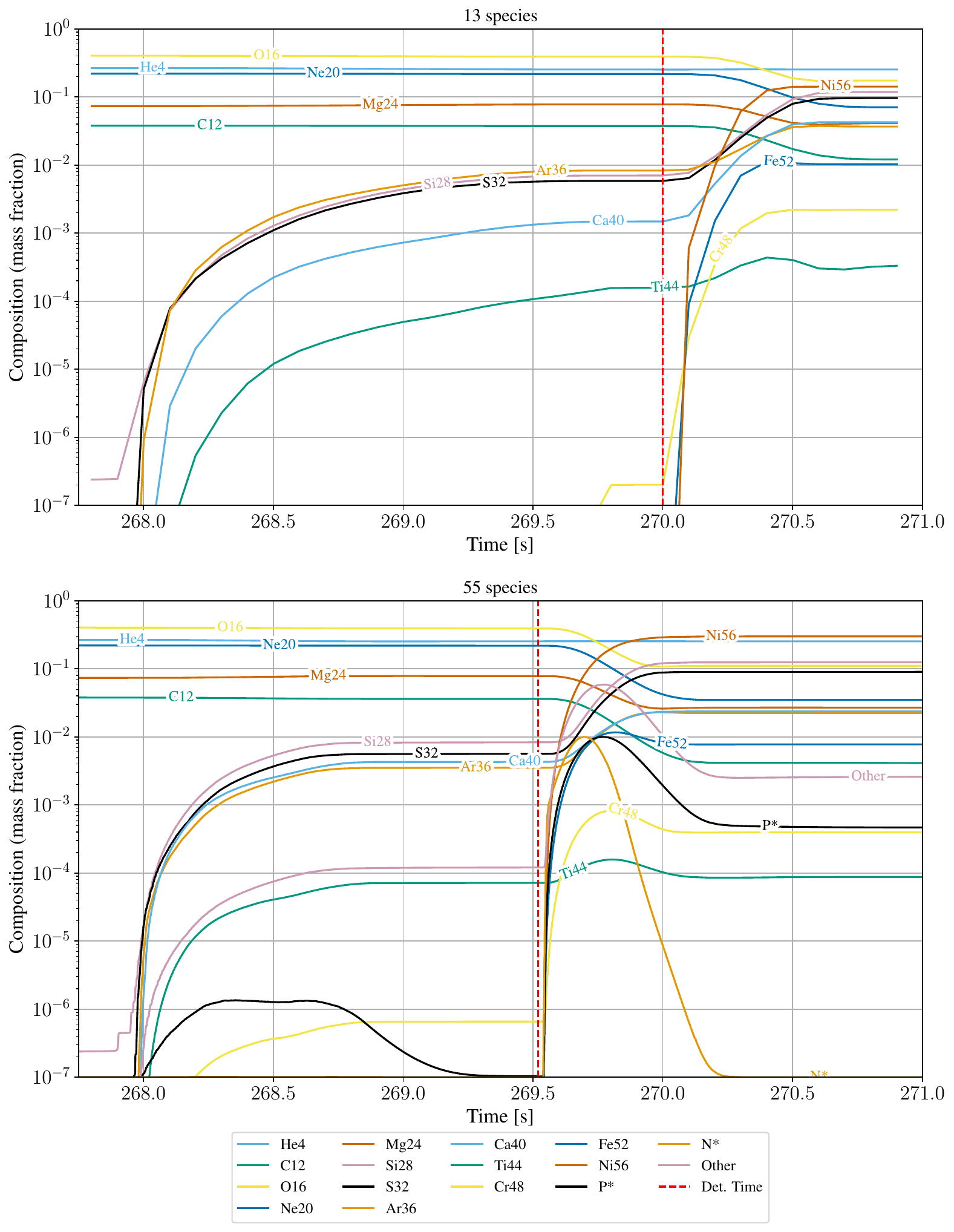}
    \caption{Comparison of time variation in composition for the simulations involving the homogenous WD structure around the explosion time using the 13-species and 55-species networks (the inspiral time is 150\,s). We can see that the time between helium detonation and convergence is shorter for the 55-species network. Note that in the right panel, the abundances of $\mathrm{p*}$ and $\mathrm{n*}$ have been scaled to appear visible on the graph, where $\mathrm{p*} = 7.296\mathrm{p} + 10^{-7}$ and $\mathrm{n*} = 1.985 \times 10^6 \mathrm{n} + 10^{-7}$.}
    \label{fig:comp_h150_13v55}
\end{figure*}

In this paper, we have considered two different nuclear networks: the 13-species and 55-species networks (see Table~\ref{tab:55_13_network_comparison}). The reactions involving He in the 13-species network are predominantly through $\alpha$ capture, contributing to the synthesis of heavier nuclei up to $^{56}$Ni, that is: 

\[
\begin{array}{rll}\label{alpha_capture}
^{4}\mathrm{He} + ^{4}\mathrm{He} & \rightarrow & ^{8}\mathrm{Be} \\
^{8}\mathrm{Be} + ^{4}\mathrm{He} & \rightarrow & ^{12}\mathrm{C} \\
^{12}\mathrm{C} + ^{4}\mathrm{He} & \rightarrow & ^{16}\mathrm{O}  \\
^{16}\mathrm{O} + ^{4}\mathrm{He} & \rightarrow & ^{20}\mathrm{Ne} \\
\cdots  \nonumber\\
^{52}\mathrm{Fe} + ^4\mathrm{He} &\rightarrow & ^{56}\mathrm{Ni}.
\end{array}
\]

while  additional $(\alpha, p)$ reactions that release protons can occur in the 55-species network. These $(\alpha, p)$ reactions can take place at somewhat lower temperatures compared to some $\alpha$ capture reactions, because they typically involve less tightly bound and lighter nuclei. Additionally, the release of a proton can be energetically favourable, often resulting in a greater energy release compared to the formation of a heavier nucleus through $\alpha$ capture. The $(\alpha, p)$ processes explain the generation of protons in the right panel of Figure \ref{fig:comp_h150_13v55}, as well as the shorter time between the onset of helium detonation and convergence. The production of neutrons in the 55-species network (see right panel of Figure \ref{fig:comp_h150_13v55}) indicates that additional reactions occur, for instance:

\begin{align*}
^{4}\mathrm{He} + ^{44}\mathrm{Ti} & \rightarrow ^{47}\mathrm{V} + n, \\
^{4}\mathrm{He} + ^{48}\mathrm{Cr} & \rightarrow ^{51}\mathrm{Mn} + n, \\
^{4}\mathrm{He} + ^{52}\mathrm{Fe} & \rightarrow ^{55}\mathrm{Co} + n, \\
^{4}\mathrm{He} + ^{56}\mathrm{Ni} & \rightarrow ^{59}\mathrm{Ni} + n.
\end{align*}

Thus, we again emphasise that the network size employed in these simulations has a critical effect on the qualitative outcomes and observables. The included reactions in the larger network allow for burning to higher atomic numbers, resulting in large differences in the quantity of relevant isotopes (e.g. $^{56}\mathrm{Ni}$). This is true even in cases where the energy release of the two networks is similar.

\bsp	
\label{lastpage}
\end{document}